\documentclass[11pt,a4paper]{article}

\usepackage{amsmath,amssymb}
\usepackage[dvipdfmx]{graphicx}
\usepackage{braket}
\usepackage{physics}
\usepackage[samesize]{cancel}
\allowdisplaybreaks[1]

\begin{document}
\title{
\begin{flushright}
\ \\*[-80pt] 
\begin{minipage}{0.2\linewidth}
\normalsize
HUPD1901 \\*[50pt]
\end{minipage}
\end{flushright}
{\Large \bf 
 Construction of RG improved effective potential in a two real scalar system
\\*[20pt]}}

\author{ 
\centerline{
Hideaki~Okane \footnote{hideaki-ookane@hiroshima-u.ac.jp}}
\\*[20pt]
\centerline{
\begin{minipage}{\linewidth}
\begin{center}
{\it \normalsize
Graduate School of Science, Hiroshima University, Higashi-Hiroshima 739-8526, Japan}
\end{center}
\end{minipage}}
\\*[50pt]}

\date{
\centerline{\small \bf Abstract}
\begin{minipage}{0.9\linewidth}
\medskip 
\medskip 
\small 
We study the improvement of effective potential by renormalization group (RG) equation in a two real scalar system.
We clarify the logarithmic structure of the effective potential in this model.
Based on the analysis of the logarithmic structure of it, we find that the RG improved effective potential up to $L$-th-to-leading log order can be calculated by the $L$-loop effective potential and $(L+1)$-loop $\beta$ and $\gamma$ functions.
To obtain the RG improved effective potential, we choose the mass eigenvalue as a renormalization scale.
If another logarithm at the renormalization scale is large, we decouple the heavy particle from the RG equation and we must modify the RG improved effective potential.
In this paper we treat such a situation and evaluate the RG improved effective potential.
Although this method was previously developed in a single scalar case, we implement the method in a two real scalar system.
The feature of this method is that the choice of the renormalization scale does't change even in a calculation of higher leading log order.
Following the our method one can derive the RG improved effective potential in a multiple scalar model.
\end{minipage}}

\begin{titlepage}
\maketitle
\thispagestyle{empty}
\end{titlepage}

\flushbottom

\section{Introduction}
Effective potential improved by renormalization group (RG) equation is widely applied in particle physics.
In refs.~\cite{EliasMiro:2011aa,Degrassi:2012ry,Bezrukov:2012sa,Alekhin:2012py,Masina:2012tz,Buttazzo:2013uya}, stability of electroweak vacuum is studied through the evaluation of the RG improved effective potential in high energy scale.
In addition, using the RG improved effective potential, authors in refs.~\cite{Coleman:1973jx,Hempfling:1996ht,Meissner:2006zh,Chang:2007ki,Foot:2007iy,Iso:2009ss,Holthausen:2009uc,AlexanderNunneley:2010nw,Ishiwata:2011aa,Holthausen:2013ota,Haba:2015qbz,Endo:2015ifa} investigate the possibility that spontaneous symmetry breaking is realized by quantum correction to the effective potential.
In this way, the RG improved effective potential is frequently utilized.
\par

There have been many researches for the RG improvement of the effective potential since a study by Coleman and Weinberg~\cite{Coleman:1973jx}.
In refs.~\cite{Kastening:1991gv,Bando:1992np,Ford:1992mv}, the RG improved effective potential in a single field is derived.
If utilizing the RG invariance of the effective potential the renormalization scale $\mu$ is set as a field dependent mass $M(\phi,\mu)$, a logarithm $\log(M(\phi,\mu)^2/\mu^2)$ become zero.
In that case, the logarithmical perturbative expansion of the effective potential including $(\log(M(\phi,\mu)^2/\mu^2))^L$ at a $L$-loop level is stable because of $\log(M(\phi,\mu)^2/\mu^2)=0$.
This is a essential point for the construction of the RG improved effective potential.
If theory includes multiple fields, the situation is not so simple.
Taking $M(\phi,\mu)$ as a renormalization scale, one cannot guarantee that the logarithm $\log(M'(\phi,\mu)^2/\mu^2)$ coming from another field is always small.
If the logarithm is large, it leads to the breakdown of the perturbative expanssion for the effective potential.
In refs.~\cite{Einhorn:1983fc,Bando:1992wy,Ford:1996hd,Ford:1996yc,Casas:1998cf}, the methods to solve such a problem are studied.
The methods are classified into two types.
In refs.~\cite{Einhorn:1983fc,Ford:1996hd,Ford:1996yc}, multiple renormalization scales are introduced and the each logarithms are suppressed by the muliple renormalization scales.
On the other hand, decoupling theorem~\cite{Appelquist:1974tg} is applied in refs.~\cite{Bando:1992wy,Casas:1998cf}.
If large logarithm appears in the calculation of the effective potential, the heavy particle is decoupled.
Since the remaining logarithm is only one of a light field, the calculation of the RG improved effective potential is the same as the way explained in the single field case.
Note that these methods are applied to theory including only a single scalar field.
\par

If multiple scalar fields are introduced, the analysis of the RG improved effective potential is complicated because the masses appearing in the logarithms depend on the multiple classical background fields such as $M(\phi_1,\phi_2)$.
The problem is addressed in refs.~\cite{Steele:2014dsa,Iso:2018aoa,Chataignier:2018aud}.
In ref.~\cite{Steele:2014dsa}, the RG improved effective potential is calculated with the introduction of the multiple renormalization scale.
In ref.~\cite{Iso:2018aoa}, which extends the method of ref.~\cite{Casas:1998cf}, a step function for the automatic decoupling of a heavy particle is introduced in the effective potential.
Moreover, effective action is analyzed to take wave function renormalization into account.
In ref.~\cite{Chataignier:2018aud}, a new method is suggested.
The guiding principle for the method is to choose the renormalization scale so that the total loop correction vanishes.
In ref.~\cite{Chataignier:2018kay} the RG improved effective potential in classical conformal theory is analyzed based on the method of ref.~\cite{Chataignier:2018aud}.
In the present paper, we also approach the problem for the RG improvement of the effective potential.
\par

In this paper, extending the method of ref.\cite{Bando:1992wy}, we construct the RG improved effective potentail in a two real scalar theory.
Since the method of ref.\cite{Bando:1992wy} is based on the analysis of the logarithmic structure of the effective potential, we derive the expression of the effective potential expanded with respect to all the logarithms appearing in a two real scalar system.
Based on the analysis of the logarithmic structure of the effective potential, we choose the field dependent mass eigenvalue as a renormalization scale so that one of the logarithms vanishes.
If another logarithm at the renormalization scale is small enough to be perturbative, the RG improved effective potential is calculated with the choice of the renormalization scale.
If the logarithm is large, we absorb the logarithm into the new parameters defined in low-energy scale and decouple the heavy particle from the theory.
Since the logarithm to be considered is only one of light particle, we can easily evaluate the RG improved effective potential.
The advantages of this method are as follows.
First, since this method is based on the logarithmic structure of the effective potential at any loop order, the choice of the renormalization scale doesn't need to be changed even in higher loop order.
Second, we can derive the RG improved effective potential without introducing multiple renormalization scales or a step fucntion for the decoupling.
Finally, we can easily implement the decoupling theorem by expanding the effective potential coming from quantum correction with respect to $\phi^2/m^2$ ($\phi^2=\phi_1^2+\phi_2^2$, $m$: decoupling scale).
\par

This paper is organized as follows:
In section~\ref{sec2}, we clarify the logarithmic structure of the effective potential and investigate the choice of the renormalization scale.
In section~\ref{sec3}, the massless theory is treated and the RG improved effective potential is calculated based on the analysis of section~\ref{sec2}.
In section~\ref{sec4}, we consider the massive model.
In this section, we face a situation in which the large logarithm occurs.
We decouple the heavy particle and construct the RG improved effective potential in the low-energy scale.
In section~\ref{sec5}, we summarize the procedure of RG improvement in multiple scalar model and discuss the application to other model.
In appendix~\ref{app1}, the $\beta$ and $\gamma$ functions at $1$-loop level is given.

\section{Logarithmic structure of effective potential and RG improvement}\label{sec2}
In this section, we clarify the logarithmic structure of the effective potential based on ref.~\cite{Bando:1992wy}.
And then, we consider the choice of the renormalization scale for the RG improvement of the effective potential.
For more specific explanation, we consider a two real scalar system as an example.
The Lagrangian is given as follwos
\begin{align}
\mathcal{L}&=\frac{1}{2}(\partial \sigma)^2+\frac{1}{2}(\partial \chi)^2-\frac{m_1^2}{2}\sigma^2-\frac{m_2^2}{2}\chi^2-\frac{\lambda_1}{4!}\sigma^4-\frac{\lambda_2}{4!}\chi^4-\frac{\lambda_3}{4}\sigma^2\chi^2-\Lambda.\label{lag}
\end{align}
We suppose that this model has $Z_2\times Z_2$ symmetry: $\sigma\rightarrow -\sigma$ and $\chi\rightarrow -\chi$.
Following ref.~\cite{Bando:1992wy}, we factor out a coupling constant $1/\lambda_1$ from the Lagrangian\footnote{In this paper, we assume that all the quartic coupling costants are comparable to each other $(\mathcal{O}(\lambda_1)\sim \mathcal{O}(\lambda_2)\sim \mathcal{O}(\lambda_3))$ and perturbative. Under the assumption, the choice of $\lambda_1$ doesn't affect the final expression eq.~\eqref{RGV}. That is to say, factoring out $\lambda_2$( or $\lambda_3$) replaced by $\lambda_1$, one obtains the same result eq.~\eqref{RGV}.},
\begin{align}
\begin{split}
\mathcal{L}&=\frac{1}{\lambda_1}\bigg(\frac{1}{2}\big\{\partial (\sqrt{\lambda_1}\sigma)\big\}^2+\frac{1}{2}\big\{\partial (\sqrt{\lambda_1}\chi)\big\}^2-\frac{m_1^2}{2}(\sqrt{\lambda_1}\sigma)^2-\frac{m_2^2}{2}(\sqrt{\lambda_1}\chi)^2\\
&\qquad-\frac{1}{4!}(\sqrt{\lambda_1}\sigma)^4-\frac{\lambda_2/\lambda_1}{4!}(\sqrt{\lambda_1}\chi)^4-\frac{\lambda_3/\lambda_1}{4}(\sqrt{\lambda_1}\sigma)^2(\sqrt{\lambda_1}\chi)^2-\lambda_1\Lambda\bigg).\label{lagg}
\end{split}
\end{align}
Next, we shift the fields ($\sigma,\chi$) by classical background fields $(\phi_1,\phi_2)$, respectively,
\begin{align}
\begin{split}\nonumber
\sigma &\rightarrow \phi_1+\sigma,\\
\chi &\rightarrow \phi_2+\chi,
\end{split}
\end{align}
and then redefine the quantum fields $\sqrt{\lambda_1}\sigma$ and $\sqrt{\lambda_1}\chi$ as $\sigma$ and $\chi$, respectively.
After the shift and the redifinition, the Lagrangian becomes
\begin{align}
\begin{split}\label{lag2}
\mathcal{L}&=\frac{1}{\lambda_1}\bigg(\frac{1}{2}(\partial \sigma)^2+\frac{1}{2}(\partial \chi)^2-\frac{M_1^2}{2}\sigma^2-\frac{M_2^2}{2}\chi^2-M_3^2\sigma\chi\\
&\qquad-\frac{x_1}{3!}\sigma^3-\frac{x_2 y_1}{3!}\chi^3-\frac{y_2}{2}(x_2\sigma+x_1\chi)\sigma\chi\\
&\qquad-\frac{1}{4!}\sigma^4-\frac{y_1}{4!}\chi^4-\frac{y_2}{4}\sigma^2\chi^2-\lambda_1V^{(0)}\bigg),
\end{split}
\end{align}
where mass parameters ($M_1^2$, $M_2^2$, $M_3^2$), cubic coupling constants ($x_1$, $x_2$) and quartic coupling constants ($y_1$, $y_2$) are introduced as follows,
\begin{align}
\begin{split}\nonumber
M_1^2&=m_1^2+\frac{\lambda_1}{2}\phi_1^2+\frac{\lambda_3}{2}\phi_2^2,\\
M_2^2&=m_2^2+\frac{\lambda_3}{2}\phi_1^2+\frac{\lambda_2}{2}\phi_2^2,\\
M_3^2&=\lambda_3\phi_1\phi_2,\\
x_1&=\sqrt{\lambda_1}\phi_1,\qquad x_2=\sqrt{\lambda_1}\phi_2,\\
y_1&=\frac{\lambda_2}{\lambda_1},\qquad\qquad y_2=\frac{\lambda_3}{\lambda_1},
\end{split}
\end{align}
and $V^{(0)}$ is a tree level effective potential,
\begin{align}
V^{(0)}&=\frac{m_1^2}{2}\phi_1^2+\frac{m_2^2}{2}\phi_2^2+\frac{\lambda_1}{4!}\phi_1^4+\frac{\lambda_2}{4!}\phi_2^4+\frac{\lambda_3}{4}\phi_1^2\phi_2^2+\Lambda.\label{treepo}
\end{align}
From the rewritten Lagrangian eq.~\eqref{lag2} and the tree potential eq.~\eqref{treepo}, we can find that the theory is described by the following parameters,
\begin{align}
\text{mass parameters}&:&& M_1^2,~ M_2^2,~ M_3^2,\label{masspara}\\
\text{cubic coupling constants}&:&& x_1,~ x_2,\\
\text{quartic coupling constants}&:&& \lambda_1,~ y_1,~ y_2,\\
\text{constant term}&:&& \Lambda.
\end{align}
Moreover, since it is inconvenient for the mass matrix not to be diagonal, we rotate the mass matrix by introducing new states ($\sigma_d$ and $\chi_d$) and mixing angle ($\theta$)
\begin{align}\nonumber
\begin{pmatrix}\sigma\\ \chi\end{pmatrix}=
\begin{pmatrix}\cos\theta&-\sin\theta\\ \sin\theta&\cos\theta\end{pmatrix}
\begin{pmatrix}\sigma_d \\ \chi_d \end{pmatrix},
\quad \tan(2\theta)=\frac{2M_3^2}{M_1^2-M_2^2},
\end{align}
and then the mass matrix is diagonalized,
\begin{align}\nonumber
\begin{pmatrix}\sigma & \chi\end{pmatrix}
\begin{pmatrix}M_1^2 & M_3^2\\
M_3^2& M_2^2\end{pmatrix}
\begin{pmatrix}\sigma\\ \chi\end{pmatrix}=
\begin{pmatrix}\sigma_d & \chi_d\end{pmatrix}
\begin{pmatrix}M_+^2 & 0\\
0& M_-^2\end{pmatrix}
\begin{pmatrix}\sigma_d\\ \chi_d\end{pmatrix},
\end{align}
where mass eigenvalues are
\begin{align}\nonumber
M_\pm^2=\frac{1}{2}\bigg(M_1^2+M_2^2\pm\sqrt{(M_1^2-M_2^2)^2+4M_3^4}\bigg).
\end{align}
For later discussion, the coordinate $(\phi_1,\phi_2)$ is translated to polar coodinate $(\phi,\beta)$,
\begin{align}
\phi^2=\phi_1^2+\phi_2^2,\qquad \tan\beta=\frac{\phi_2}{\phi_1}.
\end{align}
From now on, the mass eigenvalues and the effective potential are written with the polar coodinate $(\phi,\beta)$.
\par

In this stage, we can replace the three mass parameters ($M_1^2$, $M_2^2$, $M_3^2$) in eq.~\eqref{masspara} by mass eigenvalues ($M_\pm^2$) and mixing angle ($\theta$).
Namely, the model is described in terms of the following parameters,
\begin{align}
\text{mass eigenvalues}&:&& M_\pm^2,\label{masseigen}\\
\text{mixing angle}&:&& \theta,\\
\text{cubic coupling constants}&:&& x_1,~ x_2,\\
\text{quartic coupling constants}&:&& \lambda_1,~ y_1,~ y_2,\\
\text{constant term}&:&& \Lambda.\label{constantterm}
\end{align}
This information is so important that using these parameters we can write down effective potential at $L$-loop level as
\begin{align}\label{Lpo}
V^{(L)}=\lambda_1^{L-1}M_-^4\bigg[\text{function of } \log\bigg(\frac{M_-^2}{\mu^2}\bigg),\log\bigg(\frac{M_+^2}{\mu^2}\bigg), P\bigg],
\end{align}
where $P$ is the generic term of $(p_1, \cdots, p_7)$,
\begin{align}
&p_1=\frac{M_+^2}{M_-^2},~ p_2=\theta,~ p_3=\frac{x_1^2}{M_-^2},~ p_4=\frac{x_2^2}{M_-^2},\\
&p_5=y_1,~ p_6=y_2,~ p_7=\lambda_1\frac{\Lambda}{M_-^4}.
\end{align}
Let us explain why $L$-loop effective potential can be written as eq.~\eqref{Lpo}.
Since $\lambda_1$ can be treated like a $\hbar$ in front of action, $L$-loop effective potential is proportional to $\lambda_1^{L-1}$.
The part of square brackets $[\cdots]$ in eq. \eqref{Lpo} are dimensionless because $M_-^4$ is extracted as dimensionful part of $V^{(L)}$.
So since we introduce dimensionless parameters $(p_1,\cdots, p_7)$ based on eqs.~\eqref{masseigen}-\eqref{constantterm}, the part of square brackets $[\cdots]$ can be written in terms of two logarithms ($\log(M_-^2/\mu^2)$ and $\log(M_+^2/\mu^2)$) and dimensionless parameters $(p_1,\cdots, p_7)$.
\par

As well known, since $L$-loop effective potential $V^{(L)}$ contains $L$-th power of the logarithm at most, one can express $V^{(L)}$ with respect to $\log(M_-^2/\mu^2)$ and $\log(M_+^2/\mu^2)$,
\begin{align}
V^{(L)}=\frac{M_-^4}{\lambda_1}\sum_{l=0}^L\sum_{k=0}^{L-l}\lambda_1^lv_{L-(k+l),k}^{(L)}(P)s_1^{L-(l+k)}s_2^k,
\end{align}
where multiplying each logarithm by $\lambda_1$ we define $s_1$ and $s_2$
\begin{align}\nonumber
s_1&=\lambda_1\log\bigg(\frac{M_-^2}{\mu^2}\bigg),\qquad
s_2=\lambda_1\log\bigg(\frac{M_+^2}{\mu^2}\bigg).
\end{align}
Finally, by summing up $V^{(L)}$ from $L=0$ to $L=\infty$, we obtain the total effective potential expressed in terms of $s_1$ and $s_2$
\begin{align}
V&=\sum_{L=0}^{\infty}V^{(L)}=\frac{M_-^4}{\lambda_1}\sum_{l=0}^\infty \lambda_1^l f_l(P, s_1, s_2),\label{Vlog}\\
f_l(P, s_1, s_2)&=\sum_{L=l}^\infty\sum_{k=0}^{L-l}v_{L-(l+k),k}^{(L)}(P)s_1^{L-(l+k)}s_2^k.\label{fl}
\end{align}
In this expression the power of $\lambda_1$ gives the order of leading log series expansion.
In this sense, $f_l$ means the $l$-th-to-leading log function of effective potential.
\par

Next, we consider the choice of renormalization scale.
As well known, the effective potential satisfies the RG equation,
\begin{align}
\mathcal{D}V=\mu\frac{d}{d\mu}V=0,
\end{align}
where RG differential operator is given as
\begin{align}\label{D}
\mathcal{D}=\mu\frac{d}{d\mu}=\mu\frac{\partial}{\partial\mu}-\sum_{X}\gamma_X X\frac{\partial}{\partial X}+\sum_{Y}\beta_Y\frac{\partial}{\partial Y},
\end{align}
where
\begin{align}\nonumber&\gamma_X=-\frac{\mu}{X}\frac{dX}{d\mu}, \qquad\qquad\beta_Y=\mu\frac{dY}{d\mu},\\
&X=m_1^2,m_2^2,\Lambda,\phi_1,\phi_2,\qquad
Y=\lambda_1,\lambda_2,\lambda_3.
\end{align}
These specific $\beta$ and $\gamma$ functions are given in appendix \ref{app1}.
And then we can obtain the solution of the RG equation as
\begin{align}\label{sol}
V(\phi,\beta,Q;\mu_0^2)=V\left(\bar{G}(t,\beta)\phi,\beta,\bar{Q}(t);\mu_0^2e^{2t}\right),
\end{align}
where we use a shorthand notation $Q(=m_1^2,m_2^2,\lambda_1,\lambda_2,\lambda_3,\Lambda)$ and introduce $t$ to express the renormalization scale $\mu^2$ as $\mu^2(t)=\mu_0^2e^{2t}$.
Also $\bar{G}(\beta,t)$ is defined as
\begin{align}
\begin{split}
\bar{\phi}_1(t)^2+\overline{\phi}_2(t)^2&=\bigg(\text{exp}\left[-2\int_0^t ds\bar{\gamma}_{\phi_1}(s)\right]\cos^2\beta+\text{exp}\left[-2\int_0^t ds\bar{\gamma}_{\phi_2}(s)\right]\sin^2\beta\bigg)\phi^2\\
&\equiv\bar{G}(\beta,t)^2\phi^2.
\end{split}
\end{align}
However because of $\gamma_{\phi_1}=\gamma_{\phi_2}=0$, from now on, we set $\bar{G}(\beta,t)=1$.
$\bar{Q}(t)$ is the solution of $\beta$ or $\gamma$ function and satisfies an initial value $Q$ at an initial renormalization scale $\mu_0^2$ or $t=0$.
The RG solution of eq.~\eqref{sol} for the effective potential means that it is independent of the renomalization scale $t$.
Since we can freely choose the renormalization scale, we look for the best choice of the renormalization scale.
Let us take the renormalization scale as follows
\begin{align}\label{mu}
\mu^2=\bar{M}_-(t)^2.
\end{align}
Since this choice leads to $\bar{s}_1(t)=0$, the RG improved effective potential expressed with eq.~\eqref{Vlog} becomes
\begin{align}\nonumber
V=\bar{M}_-(t)^4\sum_{l=0}^\infty \bar{\lambda}_1(t)^{l-1} f_l(\bar{P}, \bar{s}_1=0, \bar{s}_2),
\end{align}
where from eq.~\eqref{fl}
\begin{align}\nonumber
f_l(\bar{P}, \bar{s}_1=0, \bar{s}_2)=\sum_{L=l}^\infty v_{0,L-l}^{(L)}(\bar{P})\bar{s}_2^{L-l}.
\end{align}
Here, if we assume $\bar{s}_2\lesssim \mathcal{O}\big(\bar{\lambda}_1\big)$,
one gets the $l$-th-to-leading log function,
\begin{align}\label{fls10}
f_l(\bar{P}, \bar{s}_1=0, \bar{s}_2)=v_{0,0}^{(l)}(\bar{P})+\mathcal{O}\left(\bar{\lambda}_1\right).
\end{align}
If $\bar{s}_2\lesssim \mathcal{O}\big(\bar{\lambda}_1\big)$ and we would like to evaluate the effective potential up to $L$-th-to-leading log order, the expression is written as
\begin{align}
V&=\bar{M}_-(t)^4\sum_{l=0}^L \bar{\lambda}_1(t)^{l-1} f_l(\bar{P}, \bar{s}_1=0, \bar{s}_2)\bigg|_{\bar{s}_2\lesssim \mathcal{O}\big(\bar{\lambda}_1\big)}\nonumber\\
&=\bar{M}_-(t)^4\sum_{l=0}^L \bar{\lambda}_1(t)^{l-1}v_{0,0}^{(l)}(\bar{P})\bigg|_{\mu^2=\bar{M}_-^2}+\mathcal{O}\left(\bar{\lambda}_1^{L}\right)\nonumber\\
&=\sum_{l=0}^{L}V^{(l)}(\phi,\beta,\bar{Q}(t);\mu_0^2e^{2t})\bigg|_{\mu^2=\bar{M}_-^2}+\mathcal{O}\left(\bar{\lambda}_1^{L}\right)\label{RGV}.
\end{align}
We notice that the term of $\mathcal{O}\left(\bar{\lambda}_1\right)$ in eq.~\eqref{fls10}
contributes to the effective potential beyond the $L$-th-to-leading log order.
Note that the RG improved effective potential is exactly correct up to $L$-th-to-leading log order only if the RG equations for the paremters are solved up to $(L+1)$ loop level.
In summary, if one prepares the $L$-loop effective potential and $(L+1)$-loop $\beta$ and $\gamma$ functions, one can construct the RG improved effective potential eq.~\eqref{RGV} up to $L$-th-to-leading log order for the case of $\bar{s}_2\lesssim \mathcal{O}\left(\bar{\lambda}_1\right)$.
\par

We comment on the variables of the effective potential.
Originally, the effective potential has the three variables $(\phi,\beta,t)$.
However now that $\mu(t)^2$ is taken to be equal to $\bar{M}_-(\phi,\beta,t)^2$, these variables are related.
In our paper, we show that we can solve $\mu(t)^2=\bar{M}_-(\phi,\beta,t)^2$ analytically with respect to $\phi$ and construct the RG improved effective potential by using the solution of $\phi$.
\footnote{Moreover, note that although the dimensionless parameters $P$ are introduced for the derivation of the logarithmic structure of the effective potential, the final expression is written in terms of the parameters $Q$.
Namely, we do not use the dimensionless parameters $P$ but the parameters $Q$ for the calculation of the RG improved effective potential eq.~\eqref{RGV} .}
\par

Since the above prescription is correct only in the case of $\bar{s}_2\lesssim \mathcal{O}\left(\bar{\lambda}_1\right)$, we must consider the method of the RG improvement for the case of $\bar{s}_2> \mathcal{O}\left(\bar{\lambda}_1\right)$.
In that case, as seen from the logarithm $\log\left(\bar{M}_+(t)^2/\bar{M}_-(t)^2\right)$ of $\bar{s}_2$,  the relative magnitude of the mass eigenvalues is large.
In such a case, we make use of decoupling theorem.
The decoupling of the heavy particle means that the logarithm of the particle is absorbed into the parameters defined in the effective theory.
The remaining logarithm is only one of the light particle.
If the theory includes only a single logarithm, by setting the renormalization scale as the light mass, the RG improved effective potential can be calculated.
We discuss more specifically the situation in section~\ref{sec4}.

\section{RG improved effective potential in two real scalar systerm (massless case)}\label{sec3}
We specifically calculate the RG improved effective potential by the method constructed in section~\ref{sec2}.
In this section we treat the two real scalar model without mass parameters.
The procedure for the construction of the RG effective potential is as follows.
Because of taking the renormalization scale as $\mu_0^2e^{2t}=\bar{M}_-(t)^2$, we solve it with respect to $\phi$.
Substituting the $\phi$ into the mass eigenvalue $M_{+}^2$ and the effective potential, we can evaluate $\log(\bar{M}_+(t)^2/\bar{M}_-(t)^2)$ and the effective potential.
If the logarithm is small enough for $\bar{s}_2$ to be the oreder of $\bar{\lambda}_1$, we can use the expression of eq.~\eqref{RGV} as the effective potential up to $L$-th-to-leading log order.
\par

In order to obtain $\phi$ with $\mu_0^2e^{2t}=\bar{M}_-(t)^2$ satisfied, we solve it in terms of $\phi$.
In present model, the mass eigenvalues $M_\pm^2$ are written as
\begin{align}
\begin{split}\nonumber
M_\pm^2&=\frac{\phi^2}{4}\bigg((\lambda_1+\lambda_3)\cos^2\beta+(\lambda_3+\lambda_2)\sin^2\beta\\
&\qquad\pm\sqrt{\big((\lambda_1-\lambda_3)\cos^2\beta+(\lambda_3-\lambda_2)\sin^2\beta\big)^2+16\lambda_3^2\sin^2\beta\cos^2\beta}\bigg)\\
&\equiv \lambda_\pm(\beta)\phi^2.
\end{split}
\end{align}
So we can easily obtain $\phi$ from $\mu_0^2e^{2t}=\bar{M}_-(t)^2$,
\begin{align}
\phi^2=\frac{\mu_0^2e^{2t}}{\bar{\lambda}_-(\beta,t)}.\label{muphi}
\end{align}
As mentioned above, now the $\phi$ is not the variable of the effective potential and  is determined by $\beta$ and $t$.
The $\phi$ appearing in the mass eigenvalue $\bar{M}_+^2$ and the effective potential is calculated with eq.~\eqref{muphi}.
\par

\begin{figure}[t]
  \centering
  \includegraphics[width=10cm]{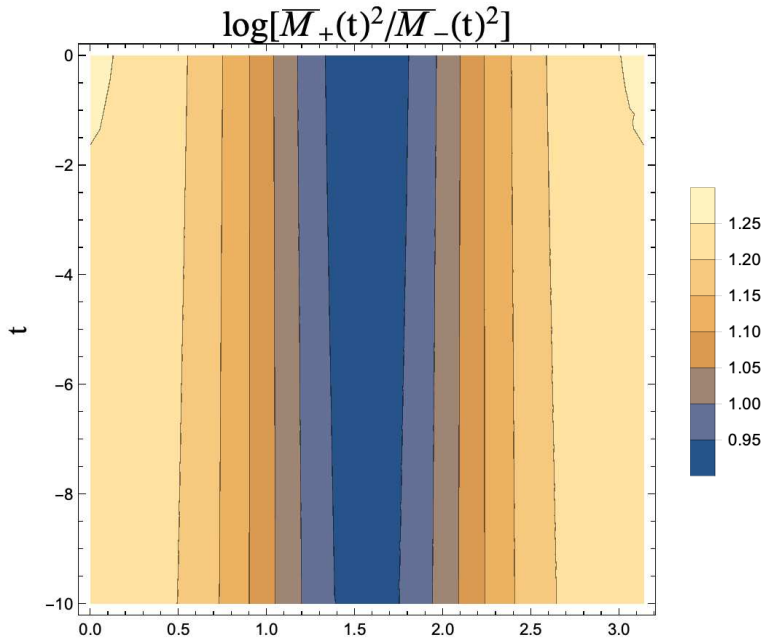}
  \vspace*{-0.5cm} 
  \caption{This figure shows a contour plot of $\log(\bar{M}_+^2/\bar{M}_-^2)$ in the regions of $\beta\in(0,\pi)$ and $t\in(-10,0)$. We take $\lambda_1=0.7$, $\lambda_2=0.5$, $\lambda_3=0.2$ and $\phi=10$ at $\mu_0^2=M_-^2$ as an initial condition.}\label{fig1}
\end{figure}

The logarithm of $\bar{s}_2$ is written as
\begin{align}
\log\bigg(\frac{\bar{M}_+(t)^2}{\mu(t)^2}\bigg)\bigg|_{\mu(t)^2=\bar{M}_-(t)^2}=\log\bigg(\frac{\bar{\lambda}_+(\beta,t)}{\bar{\lambda}_-(\beta,t)}\bigg),
\end{align}
where $\phi$ is canceled out because of the massless model.
In this stage we assign initial values of ($\lambda_1$, $\lambda_2$, $\lambda_3$) for peforming the numerical calculation.
Taking $\lambda_1=0.7$, $\lambda_2=0.5$, $\lambda_3=0.2$ and $\phi=10$ at $\mu_0^2=M_-^2$, we calculate $\log(\bar{M}_+(t)^2/\bar{M}_-(t)^2)$ for the range of $\beta\in(0,\pi)$ and $t\in(-10,0)$ by $1$-loop $\beta$ functions in figure \ref{fig1}.
In figure \ref{fig1} we see $\log(\bar{M}_+(t)^2/\bar{M}_-(t)^2)\approx 1$ in the regions of $(\beta,t)$.
Thus since we can conclude $\bar{s}_2\approx \bar{\lambda}_1$, eq.~\eqref{RGV} can be used as the RG improved effective potential.
Using the tree level effective potential and the $1$-loop $\beta$ function, the RG improved effective potential at the leading log order is given as
\begin{align}
V=\bigg(\frac{\bar{\lambda}_1(t)}{4!}\cos^4\beta+\frac{\bar{\lambda}_2(t)}{4!}\sin^4\beta+\frac{\bar{\lambda}_3(t)}{4}\sin^2\beta\cos^2\beta\bigg)\phi^4
\quad \text{with}\quad \phi^2=\frac{\mu_0^2e^{2t}}{\bar{\lambda}_-(\beta,t)},\label{RGVleading}
\end{align}
where the condition for $\phi^2$ originates from the choice of the renormalization scale $\mu^2=\bar{M}_-(t)^2$, as seen in eq.~\eqref{muphi}.
Clearly the RG improved effective potential is determined by $\beta$ and $t$.
\begin{figure}[t]
  \centering
  \includegraphics[width=8cm]{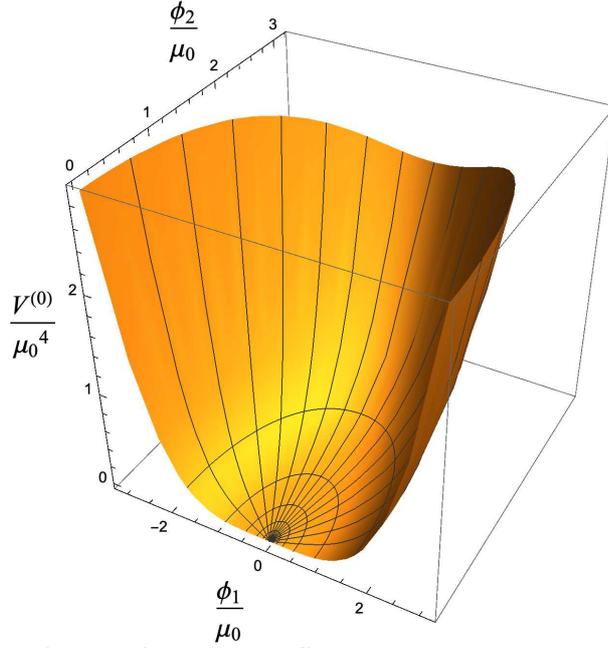}
  \vspace*{-0.5cm} 
  \caption{This is a 3D plot of the RG improve effective potential at the leading log order divided by the initial renormalization scale $\mu_0^4$ as axes of $(\phi_1/\mu_0,\phi_2/\mu_0)$. This is ploted in the regions of $\beta\in(0,\pi)$ and $t\in(-10,0)$. The initial condition is the same as figure \ref{fig1}.}\label{fig2}
\end{figure}
In figure \ref{fig2}, the RG improved effective potential is ploted as axes of $(\phi_1/\mu_0,\phi_2/\mu_0)$ for the regions of $\beta\in(0,\pi)$ and $t\in(-10,0)$.
\par

\section{RG improved effective potential in two real scalar system (massive case)}\label{sec4}
In this section we consider the massive theory in a two real scalar model.
In particular, we treat the effective potential causing spontaneous symmetry breaking.
The procedure for the construction of the RG improved effective potential is the same as the previous way.
We solve the eq.~\eqref{mu} for $\phi$ in the massive case.
In this case, because of mass parameters, the equation is a little complicated but it can be analytically solved.
Eq.~\eqref{mu} is written as follows
\begin{align}
A=\sqrt{B},
\end{align}
where
\begin{align}
\begin{split}\nonumber
A&=m_1^2+m_2^2-2\mu^2+\frac{1}{2}\bigg((\lambda_1+\lambda_3)\cos^2\beta+(\lambda_2+\lambda_3)\sin^2\beta\bigg)\phi^2,\\
B&=\bigg\{m_1^2-m_2^2+\frac{1}{2}\bigg((\lambda_1-\lambda_3)\cos^2\beta+(\lambda_3-\lambda_2)\sin^2\beta\bigg)\phi^2\bigg\}^2+4\lambda_3^2\sin^2\beta\cos^2\beta \phi^4.
\end{split}
\end{align}
Squaring the both side of $A=\sqrt{B}$, a quadratic equation for $\phi^2$ is given as,
\begin{align}
a\phi^4+2b\phi^2+c=0,
\end{align}
where
\begin{align}
\begin{split}\nonumber
a&=\lambda_1\lambda_3\cos^4\beta+\lambda_2\lambda_3\sin^4\beta+(\lambda_1\lambda_2-3\lambda_3^2)\sin^2\beta\cos^2\beta,\\
b&=(\lambda_3\cos^2\beta+\lambda_2\sin^2\beta)(m_1^2-\mu^2)+(\lambda_1\cos^2\beta+\lambda_3\sin^2\beta)(m_2^2-\mu^2),\\
c&=4(m_1^2-\mu^2)(m_2^2-\mu^2).
\end{split}
\end{align}
We can obtain the solution $\phi^2$ as
\begin{align}
\phi^2=\frac{-b\pm\sqrt{b^2-ac}}{a}.
\end{align}
Since we solve the quadratic equation, there are two solutions for $\phi^2$.
But because the original equation is $A=\sqrt{B}$, the solution satisfies the following conditons,
\begin{align}
A>0 \quad\text{and}\quad B>0.
\end{align}
Although it is diffuclt to analytically prove whether the either solution satisfies the condition or not, using the initial values input in next subsections we confirm numerically the following results,
\begin{align}
\phi^2&=\frac{-b+\sqrt{b^2-ac}}{a}\quad\rightarrow\quad A>0 \quad\text{and}\quad B>0,\nonumber\\
\phi^2&=\frac{-b-\sqrt{b^2-ac}}{a}\quad\rightarrow\quad A<0 \quad\text{and}\quad B<0\nonumber.
\end{align}
Therefore we adopt the solution $\phi$ as
\begin{align}
\phi^2=\frac{-b+\sqrt{b^2-ac}}{a}.
\end{align}
\par

Since we get the solution $\phi$ for eq.~\eqref{mu}, we can construct the RG improved effective potential.
The expression is provided at a leading log order as
\begin{align}
\begin{split}\label{RGV2}
V&=\frac{1}{2}(\bar{m}_1(t)^2\cos^2\beta+\bar{m}_2(t)^2\sin^2\beta)\phi^2\\
&\qquad\qquad+\frac{1}{4!}(\bar{\lambda}_1(t)\cos^4\beta+\bar{\lambda}_2(t)\sin^4\beta+6\bar{\lambda}_3(t)\sin^2\beta \cos^2\beta)\phi^4+\bar{\Lambda}(t)\\
&\qquad\qquad\qquad \text{with}\qquad \phi^2=\frac{-\bar{b}(\beta,t)+\sqrt{\bar{b}(\beta,t)^2-\bar{a}(\beta,t)\bar{c}(t)}}{\bar{a}(\beta,t)}.
\end{split}
\end{align}
In the following subsections, we consider two situations for inputting the initial value of the renormalization scale.
First, taking $(m_1^2<0,m_2^2<0, -m_1^2\sim -m_2^2)$ as the mass parameters, we set  the initial renormalization scale on the vacuum which is determined by stationary condition of effective potential.
Increasing the renormalization scale from the low-energy scale at the vacuum, we analyze behavior of the RG improved effective potential in high-energy region.
Second, we input the initial values of parameters at a high-energy scale and decrease the renormalization sclale into the low-energy scale.
Assuming $(m_1^2<0,m_2^2>0, -m_1^2\ll m_2^2)$ for the mass parameters, we investigate the RG improved effective potential in the low-energy region.
As the renormalization scale decrease, the mass eigenvalue $\bar{M}_-^2$ also declines and reaches $m_2^2$ at a scale.
Since $\bar{M}_-^2$ continues to decline below the scale, we find the logarithm of $\bar{s}_2$ large.
In order to avoid the breakdown of the logarithmic perturbation, we utilize the decoupling theorem.
Applying the decoupling theorem, we derive the RG improved effective potential in the low-energy scale and visualize the behavior including the minimum value of the RG improved effective potential.

\subsection{$-m_2^2 \sim -m_1^2$}
Since we set the initial condition on the vacuum in this subsection, we derive the stationary condition for the effective potential.
Introducing a convenient notation for mass parameter and quartic coupling constant,
\begin{align}
\begin{split}\nonumber
m(\beta)^2&=m_1^2\cos^2\beta+m_2^2\sin^2\beta,\\
\lambda(\beta)&=\lambda_1\cos^4\beta+\lambda_2\sin^4\beta+6\lambda_3\sin^2\beta\cos^2\beta,\\
\end{split}
\end{align}
we can write the effective potential at a tree level,
\begin{align}\nonumber
V^{(0)}=\frac{m(\beta)^2}{2}\phi^2+\frac{\lambda(\beta)}{4!}\phi^4+\Lambda.
\end{align}
We calculate the stationary conditions for the effective potential,
\begin{align}
\frac{\partial V^{(0)}}{\partial \phi}=0,\quad \frac{\partial V^{(0)}}{\partial \beta}=0.
\end{align}
From $\frac{\partial V^{(0)}}{\partial \phi}=0$, we derive the following condition,
\begin{align}
\phi^2=-6\frac{m(\beta)^2}{\lambda(\beta)}.\label{phi2}
\end{align}
Combining this condition and $\frac{\partial V^{(0)}}{\partial \beta}=0$, we get the stationary condition for $\beta$,
\begin{align}\label{beta}
\beta=\arccos\bigg[\frac{\lambda_2m_1^2-3\lambda_3m_2^2}{(\lambda_2-3\lambda_3)m_1^2+(\lambda_1-3\lambda_3)m_2^2}\bigg].
\end{align}
Substituting this $\beta$ for eq.~\eqref{phi2}, we can obtain $\phi$ on the stationary point,
\begin{align}\label{phi}
\phi^2=-6\frac{(\lambda_2-3\lambda_3)m_1^2+(\lambda_1-3\lambda_3)m_2^2}{\lambda_1\lambda_2-9\lambda_3^2}.
\end{align}
Using eqs.~\eqref{phi}-\eqref{beta}, we can calculate the vacuum expectation value and also estimate the initial renormalization scale $\mu_0^2=\bar{M}_-(t=0)^2=M_-^2$.
For simplicity in this section we impose $\Lambda=0$ at the initial point.
\par

\begin{figure}[t]
  \centering
  \includegraphics[width=10cm]{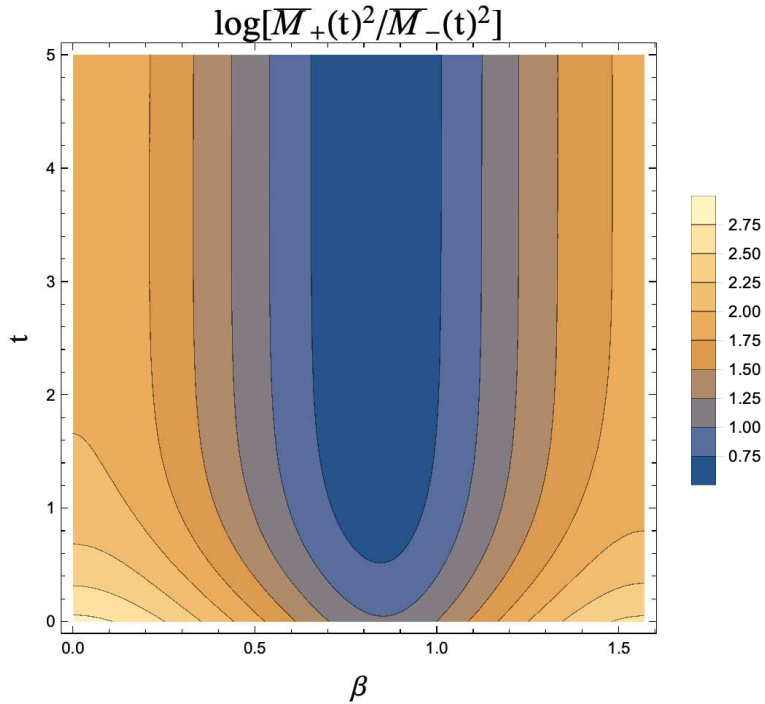} 
  \caption{The logarithm of the ratio of $\bar{M}_+(t)^2$ to $\bar{M}_-(t)^2$ is plotted in the ranges of $\beta\in(0,\frac{\pi}{2})$ and $t\in(0,5)$. The result is produced by taking $\lambda_1=0.7$, $\lambda_2=0.5$, $\lambda_3=0.1$, $m_1^2=-(160~\text{GeV})^2$ and $m_2^2=-(170~\text{GeV})^2$ as an initial condition for the RG eqution.}\label{fig3}
\end{figure}

\begin{figure}[t]
  \centering
  \includegraphics[width=7.5cm]{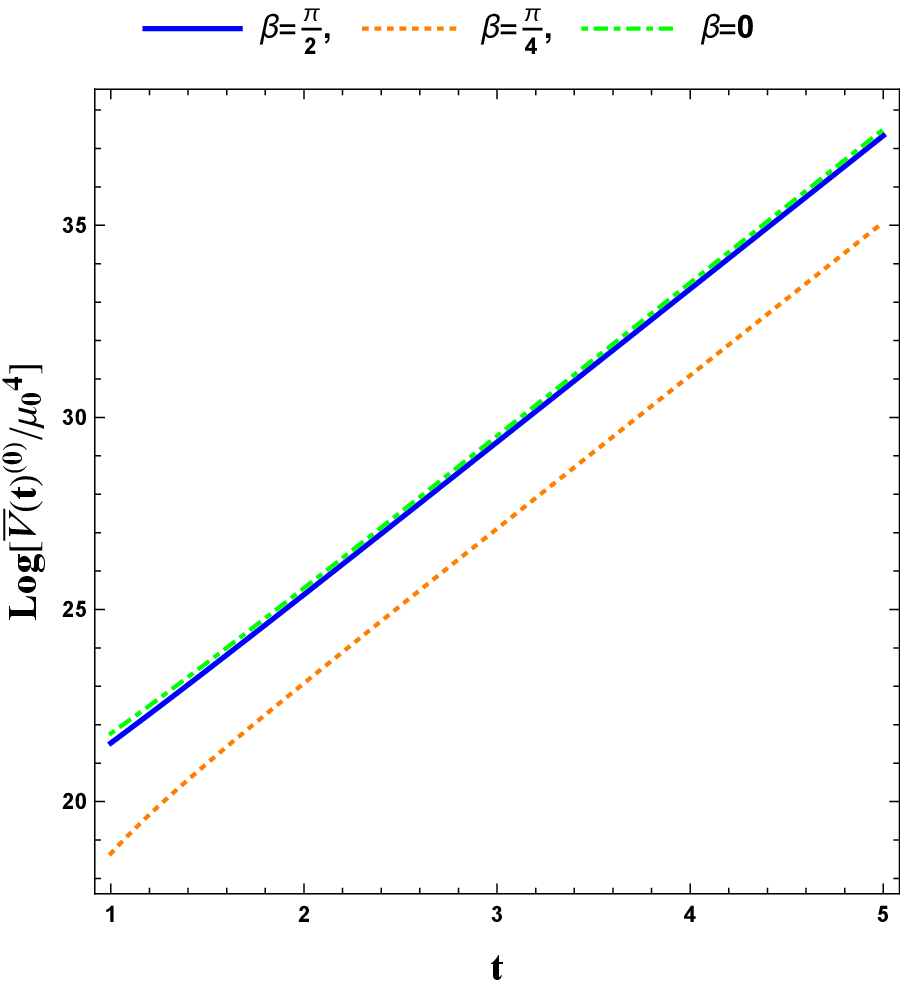}\hfill
  \includegraphics[width=7.5cm]{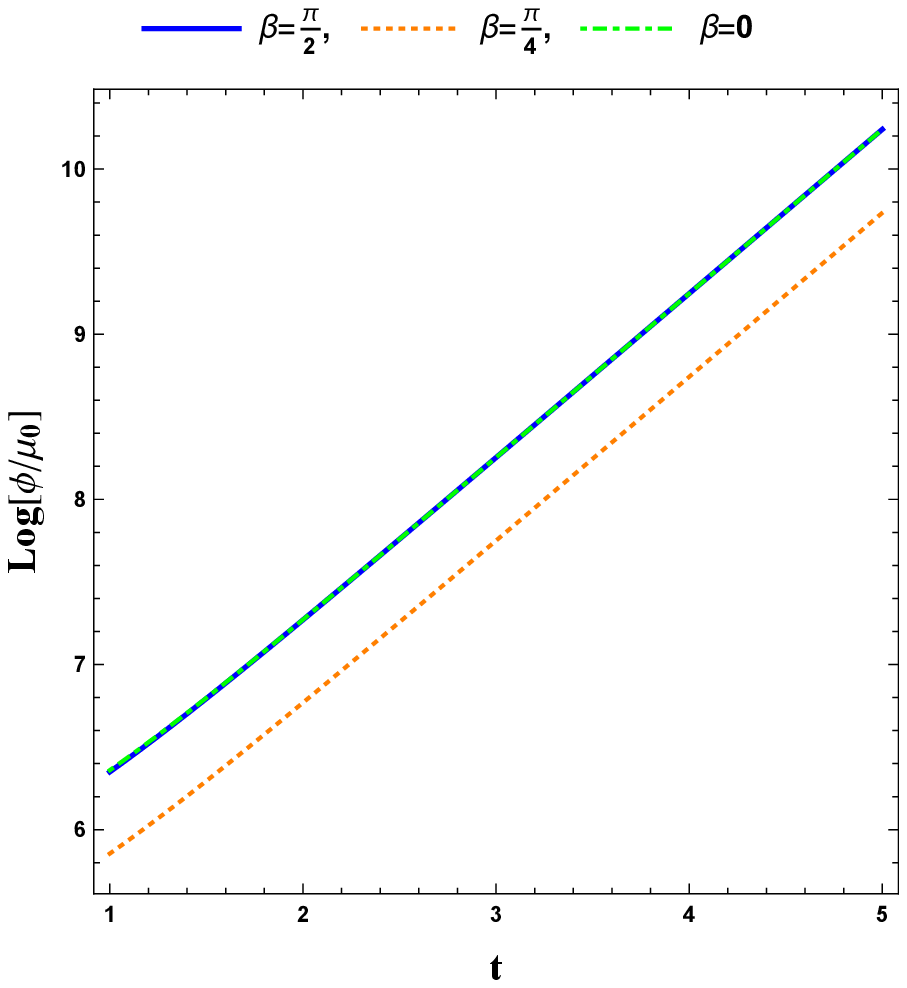}
  \caption{Left: The (dot-dash, green), (dot, orange) and (solid,blue) lines correspond to the RG improved effective potential in $\beta=0$, $\beta=\frac{\pi}{4}$ and $\beta=\frac{\pi}{2}$, respectively. Right: The (dot-dash, green), (dot, orange) and (solid,blue) lines 
correspond to $\phi$ in $\beta=0$, $\beta=\frac{\pi}{4}$ and $\beta=\frac{\pi}{2}$, respectively. The initial condition for the RG equation is the same as one in figure \ref{fig3}.}\label{fig4}
\end{figure}

Taking [$\lambda_1=0.7$, $\lambda_2=0.5$, $\lambda_3=0.1$, $m_1^2=-(160~\text{GeV})^2$ and $m_2^2=-(170~\text{GeV})^2$] as an initial condition, we get the vacuum expectation value $(\phi,\beta)=(591~\text{GeV}, 0.94)$, the initial renormalization scale $\mu_0=M_-=135\text{ GeV}$ and the mass eigenvalue $M_+=236\text{ GeV}.$
We regard the vacuum expectation value and the initial renormalization scale as a start point for the RG improved effective potential and the running parameters.
Then, we run $\log(\bar{M}_+(t)^2/\bar{M}_-(t)^2)$ by the RG equations in the regions of $t\in(0,5)$ and $\beta\in(0,\frac{\pi}{2})$.
Figure~\ref{fig3} shows the result of the logarithm.
On $\beta=0\text{ and }\beta=\frac{\pi}{2}$ in figure \ref{fig3}, the logarithm takes $2\sim 3$ in the range of $t=(0,2)$ and less than 2 for $t>2$.
In $\beta=\frac{\pi}{4}$, the logarithm is less than 1 for the all scale of $t$.
If the magnitude of the logarithm as $\log(\bar{M}_-^2/\bar{M}_+^2)\lesssim 3$ is accepted in the context of a logaritmical perturbative expansion, the RG improved effective potential is calculated with eq.~\eqref{RGV2}.
The result is shown in figure \ref{fig4}.
In the left panel of figure \ref{fig4}, the (dot-dash, green), (dot, orange) and (solid,blue) lines correspond to the RG improved effective potential in $\beta=0$, $\beta=\frac{\pi}{4}$ and $\beta=\frac{\pi}{2}$, respectively.
In the right panel of figure \ref{fig4}, the (dot-dash, green), (dot, orange) and (solid,blue) lines 
correspond to $\phi$ in $\beta=0$, $\beta=\frac{\pi}{4}$ and $\beta=\frac{\pi}{2}$, respectively ($\phi$ in $\beta=0\text{ and }\frac{\pi}{2}$ are $\phi_1$ and $\phi_2$, respectively).
In both panel of figure \ref{fig4}, the lines at $\beta=0\text{ and }\frac{\pi}{2}$ overlap to each other.
\par

We comment on the more complete discussion for the logarthmical perturbative expansion.
As explained above, there are the regions in which the logarithm is beyond $1$.
If the logarithm is considered to be large, the heavy field with mass $\bar{M}_+$ should be decoupled from the theory.
Due to this decoupling, the remaining logarithm is only $\log(\bar{M}_-^2/\mu^2)$.
Since the single logarithm can be suppressed by using the degree of freedom of the renormalization scale $\mu$, the logarithmic perturbation is stable.
Such a procedure is explained in next subsection.

\subsection{$m_2^2 \gg -m_1^2$}
\begin{figure}[t]
  \centering
  \includegraphics[width=8cm]{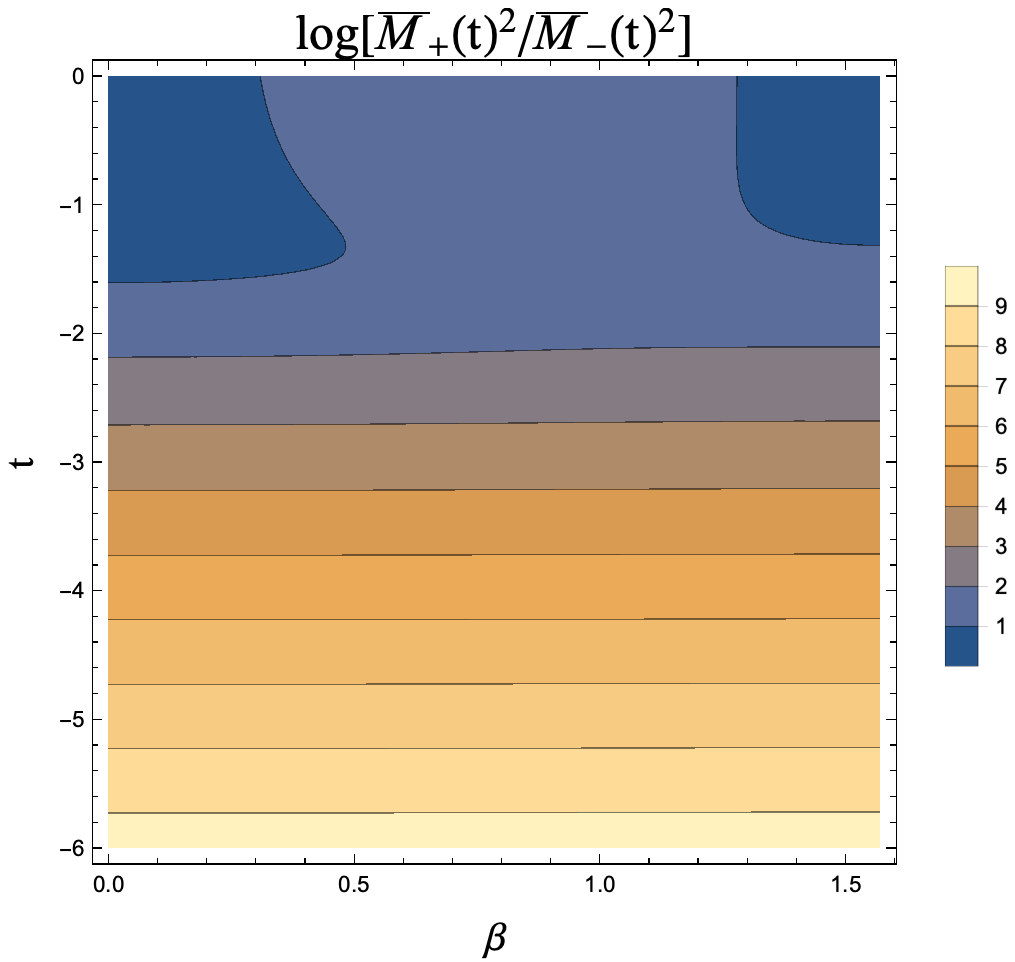}
  \caption{The logarithm of the ratio of $\bar{M}_+(t)^2$ to $\bar{M}_-(t)^2$ is plotted in the regions of $\beta\in(0,\frac{\pi}{2})$ and $t\in(-6,0)$. The initial condition is given as $\lambda_1=0.7, \lambda_2=0.6, \lambda_3=0.4,m_1^2=-(200\text{ GeV})^2, m_2^2=(3000\text{ GeV})^2, \Lambda=0\text{ at } (\phi,\beta)=(40000\text{ GeV},\frac{\pi}{4})$.}\label{fig5}
\end{figure}

\begin{figure}[t]
  \centering
  \includegraphics[width=7.5cm]{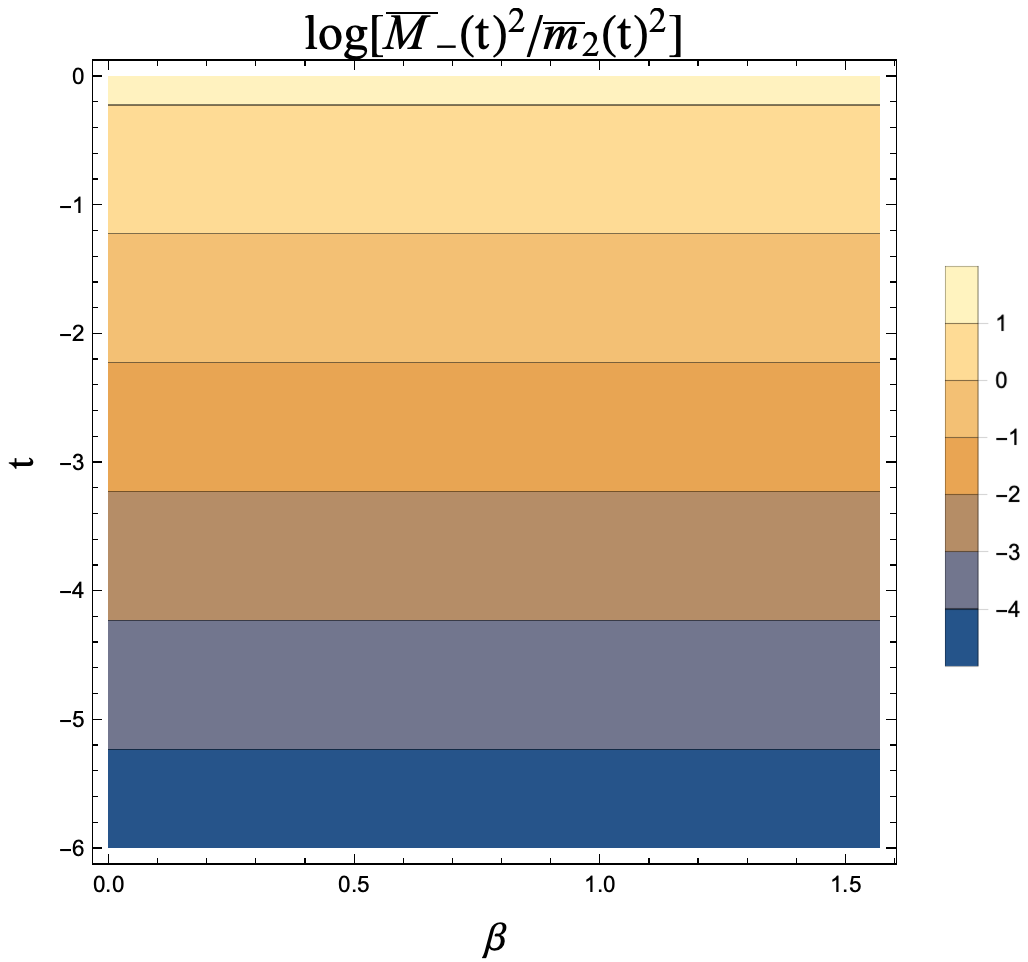}
  \includegraphics[width=7.5cm]{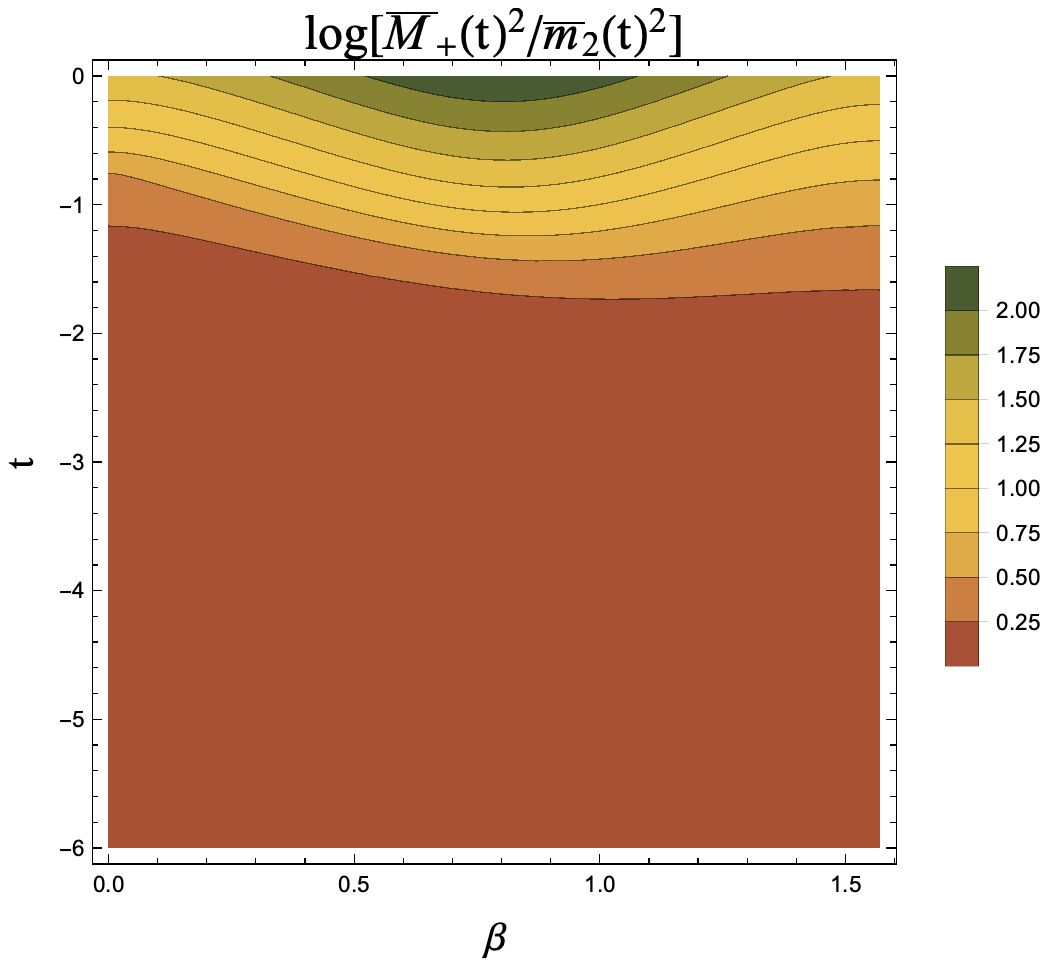}
  \caption{Left: The logarithm of the ratio of $\bar{M}_-(t)^2$ to $\bar{m}_2(t)^2$ is evaluated in the regions of $\beta\in(0,\frac{\pi}{2})$ and $t\in(-6,0)$. Right: The logarithm of the ratio of $\bar{M}_+(t)^2$ to $\bar{m}_2(t)^2$ is calculated in the regions of $\beta\in(0,\frac{\pi}{2})$ and $t\in(-6,0)$.}\label{fig6}
\end{figure}

In this subsection we impose the initial conditon at a high-energy scale and gradually decrease the renormalization scale to a scale around $-m_1^2$.
Also we suppose $m_2^2\gg -m_1^2>0$.
Setting the following the initial condition,
\begin{align}
\begin{split}\nonumber
&\lambda_1=0.7,~ \lambda_2=0.6,~ \lambda_3=0.4,\\
&m_1^2=-(200\text{ GeV})^2,~ m_2^2=(3000\text{ GeV})^2, \Lambda=0,
\end{split}
\end{align}
at $(\phi,\beta)=(40000\text{ GeV},\frac{\pi}{4})$, we evaluate the logarithm of the ratio of $\bar{M}_+(t)^2$ to $\bar{M}_-(t)^2$ in figure \ref{fig5}.
Clearly, the logarithm becomes large as the renormalization scale decreases to the low-energy scale.
This indicates the breakdown of the logarithmical perturbative expansion in the low-energy region.
For more detail, we evaluate the ratio of $\bar{M}_-(t)^2$ to $\bar{m}_2(t)^2$ in the left panel of figure \ref{fig6}.
As seen from the left panel in figure \ref{fig6}, $\bar{M}_-(t)$ steadily falls with the decreasing renormalziation scale $t$.
The ratio of $\bar{M}_+(t)^2$ to $m_2^2$ is calculated in the right panel of figure \ref{fig6}.
In contrast to  the figure in the left, the figure shows that the value of $\bar{M}_+(t)$ is comparable to $\bar{m}_2(t)$ below $t=-1$.
Therefore in figure \ref{fig6} we find out that the ratio of $\bar{M}_+(t)^2$ to $\bar{M}_-(t)^2$ increases with lower renormalization scale because $\bar{M}_-(t)$ is smaller than $\bar{m}_2(t)$ while $\bar{M}_+(t)$ is camparable to $\bar{m}_2(t)$.
\par

In order to avoid the large logarithm, we should modify the RG improved effective potential for the low-energy scale.
The way to modify the RG improvement is to utilize the decoupling theorem.
In the present case, since $\bar{M}_+(t)$ is heavier than $\bar{M}_-(t)$, the field with the mass $\bar{M}_+(t)$ should be decoupled.
Moreover, as seen in the right panel of figure \ref{fig6}, since $\bar{M}_+(t)$ is comparable to $\bar{m}_2(t)$, we factor out $\bar{m}_2(t)^2$ from the expression of $\bar{M}_+(t)^2$.
Hereafter we omit the bar of the parameters to reduce the botheration.
To implement it, we expand $M_+^2$ with respect to $\frac{\phi^2}{m_2^2}$,
\begin{align}
\begin{split}\nonumber
M_+^2&=m_2^2(1+\Delta),\\
\Delta&=\bigg(\frac{\lambda_3}{2}\cos^2\beta+\frac{\lambda_2}{2}\sin^2\beta\bigg)\frac{\phi^2}{m_2^2}+\lambda_3^2\sin^2\beta\cos^2\beta\bigg(\frac{\phi^4}{m_2^4}\bigg).
\end{split}
\end{align}
Additionally we expand the $1$-loop effective potential with $M_+^2$ in terms of $\frac{\phi^2}{m_2^2}$,
\begin{align}
\begin{split}
V_+^{(1)}&=\frac{M_+^4}{64\pi^2}\bigg(\log\bigg(\frac{M_+^2}{\mu^2}\bigg)-\frac{3}{2}\bigg)\\
&=\frac{m_2^4}{64\pi^2}\bigg(\log\bigg(\frac{m_2^2}{\mu^2}\bigg)-\frac{3}{2}\bigg)+\frac{m_2^2}{64\pi^2}(\lambda_3\cos^2\beta+\lambda_2\sin^2\beta)\bigg(\log\bigg(\frac{m_2^2}{\mu^2}\bigg)-1\bigg)\phi^2\\
&+\frac{1}{64\pi^2}\bigg\{2\lambda_3^2\sin^2\beta\cos^2\beta\bigg(\log\bigg(\frac{m_2^2}{\mu^2}\bigg)-1\bigg)+\frac{1}{4}\bigg(\lambda_3\cos^2\beta+\lambda_2\sin^2\beta\bigg)^2\log\bigg(\frac{m_2^2}{\mu^2}\bigg)\bigg\}\phi^4\\
&+\mathcal{O}\left(\frac{\phi^6}{m_2^2}\right)
\end{split}
\end{align}
In this expression we see that $\log(m_2^2/\mu^2)$ leads to the large logarithm which is not suppressed with the choice of $\mu^2=M_-^2$.
The concept of the decoupling theorem is to absorb the large logarithm into new parameters by the redefinition of the parameters.
Hence we combine the $1$-loop effective potential with the tree effective potential and redefine the new parametes to renormalize the large logarithm,
\begin{align}
\begin{split}
V^{(0)}+V_+^{(1)}&=\frac{\phi^2}{2}(\tilde{m}_1^2\cos^2\beta+\tilde{m}_2^2\sin^2\beta)\\&\qquad+\frac{\phi^4}{4!}(\tilde{\lambda}_1\cos^4\beta+\tilde{\lambda}_2\sin^4\beta+6\tilde{\lambda}_3\sin^2\beta\cos^2\beta)+\tilde{\Lambda},
\end{split}
\end{align}
where
\begin{align}
\tilde{m}_1^2&=m_1^2+\frac{\lambda_3m_2^2}{32\pi^2}\bigg(\log\bigg(\frac{m_2^2}{\mu^2}\bigg)-1\bigg)\label{repm1},\\
\tilde{m}_2^2&=m_2^2+\frac{\lambda_2m_2^2}{32\pi^2}\bigg(\log\bigg(\frac{m_2^2}{\mu^2}\bigg)-1\bigg),\\
\tilde{\lambda}_1&=\lambda_1+\frac{3\lambda_3^2}{32\pi^2}\log\bigg(\frac{m_2^2}{\mu^2}\bigg),\\
\tilde{\lambda}_2&=\lambda_2+\frac{3\lambda_2^2}{32\pi^2}\log\bigg(\frac{m_2^2}{\mu^2}\bigg),\\
\tilde{\lambda}_3&=\lambda_3+\frac{\lambda_3^2}{8\pi^2}\bigg(\log\bigg(\frac{m_2^2}{\mu^2}\bigg)-1\bigg)+\frac{\lambda_2\lambda_3}{32\pi^2}\log\bigg(\frac{m_2^2}{\mu^2}\bigg),\\
\tilde{\Lambda}&=\Lambda+\frac{m_2^4}{64\pi^2}\bigg(\log\bigg(\frac{m_2^2}{\mu^2}\bigg)-\frac{3}{2}\bigg).
\end{align}
Note that because there is no the contribution to the wave function renormalization in this model, the classical background fields don't change,
\begin{align}
\tilde{\phi}_1=\phi_1,\qquad \tilde{\phi}_2=\phi_2.\label{repp}
\end{align}
Since we use the parameters in the low-energy effective theory below $\mu^2=m_2^2$, we derive the $\beta$ and $\gamma$ functions for the redefined parameters.
To derive them, the RG differential operator in eq.~\eqref{D} is rewritten in terms of the new parameters,
\begin{align}
\begin{split}
\mathcal{D}=\mu\frac{d}{d\mu}&=(\mathcal{D}\mu)\frac{\partial}{\partial\mu}+\sum_{\tilde{X}}(\mathcal{D}\tilde{X})\frac{\partial}{\partial \tilde{X}}+\sum_{\tilde{Y}}(\mathcal{D}\tilde{Y})\frac{\partial}{\partial \tilde{Y}},\label{tildeD}\\
&=\mu\frac{\partial}{\partial\mu}-\sum_{\tilde{X}}\gamma_{\tilde{X}} \tilde{X}\frac{\partial}{\partial \tilde{X}}+\sum_{\tilde{Y}}\beta_{\tilde{Y}}\frac{\partial}{\partial \tilde{Y}},
\end{split}
\end{align}
where
\begin{align}
\tilde{X}=\tilde{m}_1^2,\tilde{m}_2^2,\tilde{\Lambda},\tilde{\phi}_1,\tilde{\phi}_2,\qquad
\tilde{Y}=\tilde{\lambda}_1,\tilde{\lambda}_2,\tilde{\lambda}_3.
\end{align}
Hence we can get the $\beta$ and $\gamma$ functions defined by the tilded parameters,
\begin{gather}
\beta_{\tilde{\lambda}_1}=\frac{3\tilde{\lambda}_1^2}{16\pi^2},
\qquad \beta_{\tilde{\lambda}_2}=\frac{3\tilde{\lambda}_3^2}{16\pi^2},
\qquad\beta_{\tilde{\lambda}_3}=\frac{\tilde{\lambda}_1\tilde{\lambda}_3}{16\pi^2},\label{bela}\\
\gamma_{\tilde{m}_1^2}=-\frac{\tilde{\lambda}_1}{16},
\qquad\gamma_{\tilde{m}_2^2}=-\frac{\tilde{\lambda}_3\tilde{m}_1^2}{16\pi^2\tilde{m}_2^2},\qquad\gamma_{\tilde{\Lambda}}=-\frac{\tilde{m}_1^4}{32\pi^2\tilde{\Lambda}}\\
\gamma_{\tilde{\phi}_1}=0,
\qquad\gamma_{\tilde{\phi}_2}=0.\label{gaph}
\end{gather}
We notice that the effect of the heavy field disappears from the RG equation in eqs.~\eqref{bela}-\eqref{gaph}.
In this sense the heavy field is decoupled from theory in the low-energy scale.
We can construct the RG improved effective potential by replacing the parameters with the tilded parameters for the effective potential in eq.~\eqref{RGV2}.
\par

Let us consider a decoupling point at which the theory is separated into the full theory and the low-energy effective theory.
From the left panel of figure \ref{fig6}, we see $\bar{M}_-(t)$ coincides with $\bar{m}_2(t)$ around $t=-1$.
Actually, as we can identify the scale as $t=-1.2$ and $\bar{M}_-(t)$ don't vary in the range of $\beta\in(0,\frac{\pi}{2})$, we use $(\beta,t)=(\frac{\pi}{2},-1.2)$ as a decoupling point.
The choice of the decoupling point is valid because the logarithm in eqs.~\eqref{repm1}-\eqref{repp} is suppressed at the scale when $\bar{M}_-(t)$ becomes equal to $\bar{m}_2(t)$.
Now we can solve the RG equations for all the paramters from the initial scale to the low-energy scale.
\begin{figure}[t]
  \centering
  \includegraphics[width=7.5cm]{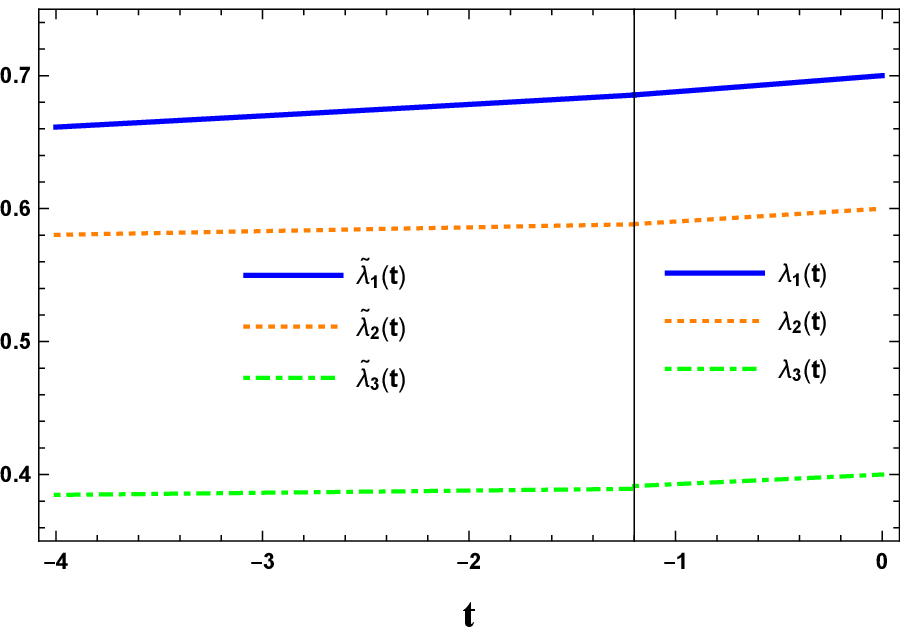}
  \includegraphics[width=7.5cm]{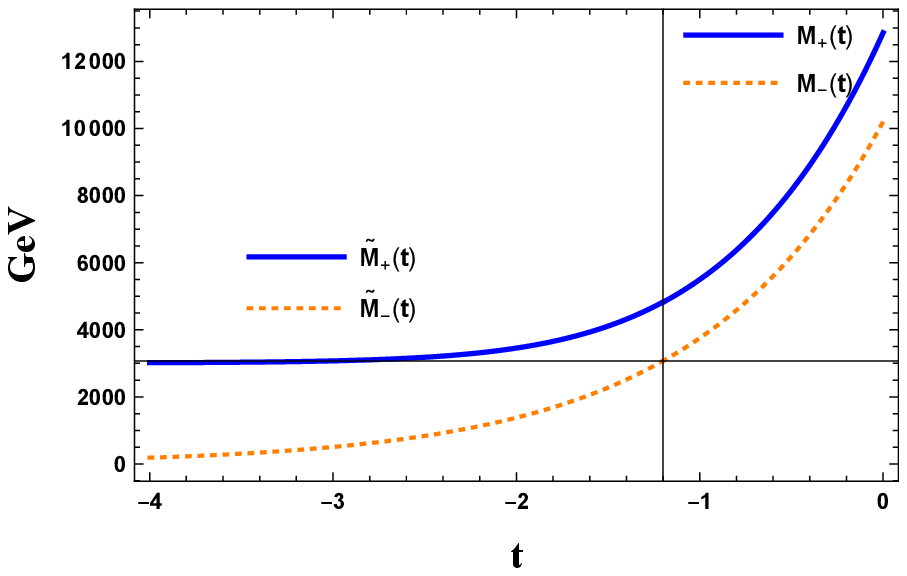}
    \caption{Left: The running of the quartic coupling constants is solved. The (solid, blue), (dot, orange) and (dot-dash, green) lines denote the running of $\lambda_1$, $\lambda_2$ and $\lambda_3$, respectively. The vertical line is the decoupling scale with $t=-1.2$. Right: The dependence of the mass eigenvalues $(M_-^2,M_+^2)$ on the renormalization scale $t$ is plotted.}\label{fig7}
\end{figure}
In the left panel of figure \ref{fig7}, the quartic coupling constants are solved from $t=0$ to $t=-4$.
We can confirm the slight threshold correction for $\bar{\lambda}_3$.
The difference between $\bar{\lambda}_3(t=-1.2)$ and $\bar{\tilde{\lambda}}_3(t=-1.2)$ normalized by $\bar{\lambda}_3(t=-1.2)$ is 0.02.
In the right panel of figure \ref{fig7}, we run the mass eigenvalues in the same range.
The $\bar{\tilde{M}}_-^2$ continues to decrease as the renormalization scale is lowered, while the $\bar{\tilde{M}}_+^2$ converges to about $3000 \text{ GeV}$.
\begin{figure}[t]
  \centering
  \includegraphics[width=7.5cm]{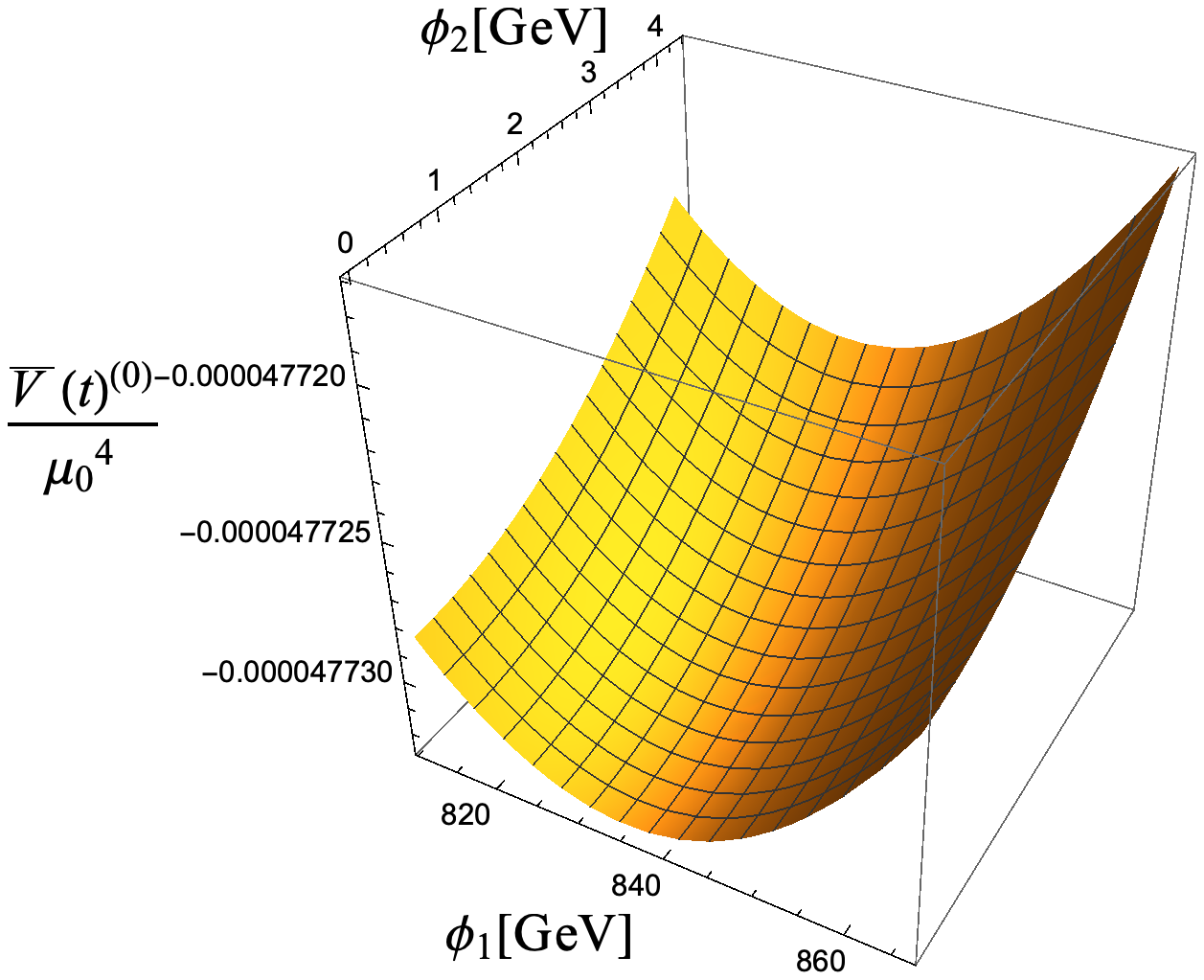}
  \includegraphics[width=7.5cm]{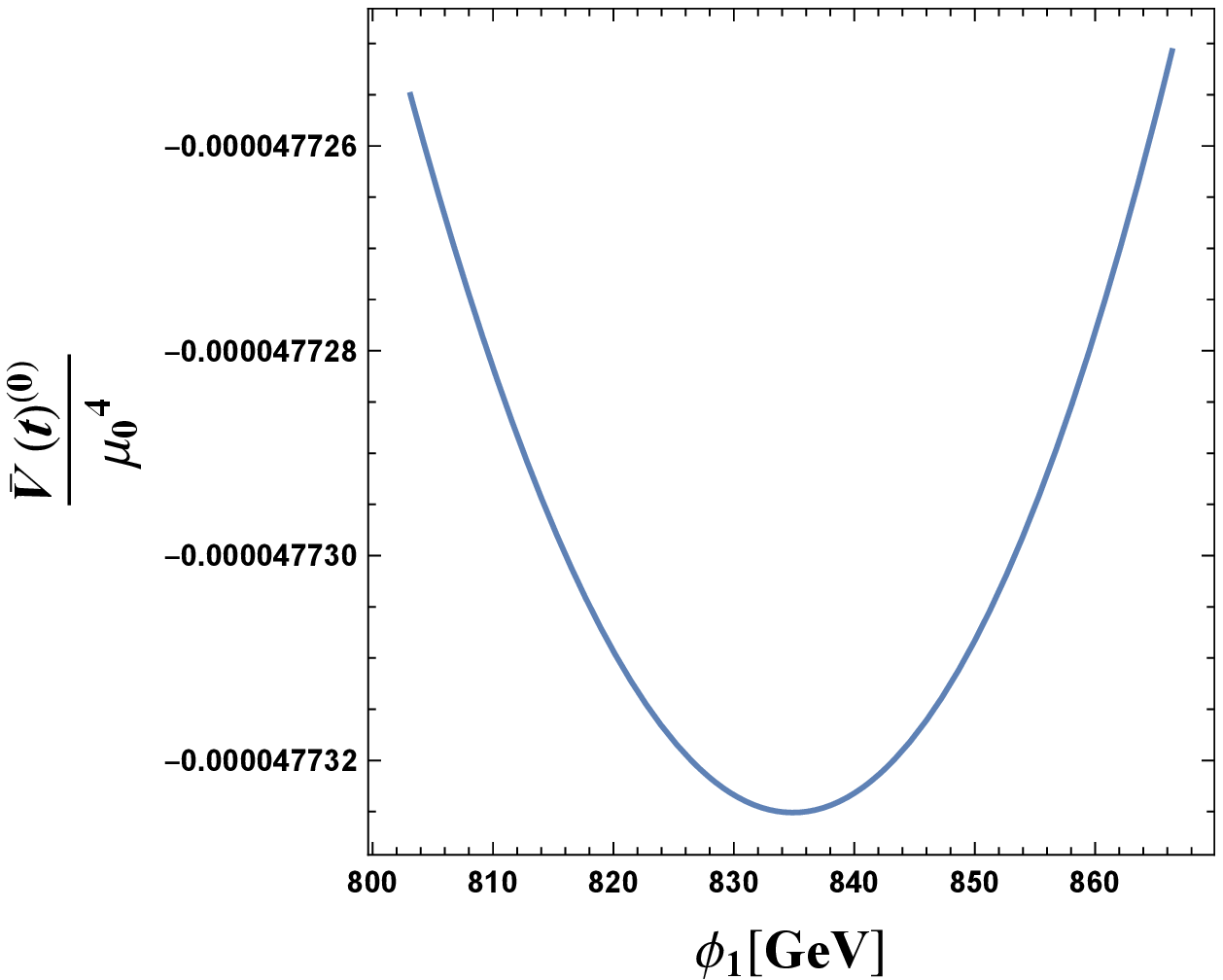}
    \caption{Left: The 3D plot of the RG improved effective potential is evaluated as a function of $(\phi_1,\phi_2)$. Right: The RG improved effective potential is plotted as a function of  $\phi_1$ with $\phi_2$ equal to zero ($\phi_2=0$). From the minimum point of the RG improved effective potential, the vacuum expectation value is estimated as ($\beta=0, t=-3.25$).}\label{fig8}
\end{figure}
In the left panel of figure \ref{fig8}, the RG improved effective potential is plotted as a function of $(\phi_1,\phi_2)$.
We can find the minimum value of the RG improved effective potential.
This point corresponds to the vacuum in the present model.
The right panel of figure \ref{fig8} shows the behavior of the RG improved effective potential as a function of $\phi_1$ with $\phi_2$ equal to zero ($\phi_2=0$).
From the evaluation of the RG improved effective potential, the vacuum expectation value correspond to $(\beta,t)=(0,-3.25)$.
Substituting them to the mass eigenvalues, we obtain the values of the masses,
\begin{align}
\bar{\tilde{M}}_-\bigg|_{\substack{\beta=0 \\ t=-3.25}}=396\text{ GeV},\qquad \bar{\tilde{M}}_+\bigg|_{\substack{\beta=0 \\ t=-3.25}}=3007\text{ GeV}.
\end{align}

\section{Summary and Discussion}\label{sec5}
In this paper we have studied the RG improvement of the effective potential in a two real scalar system.
In section~\ref{sec2} we clarify the logarithmic structure of the effective potential.
If we choose $\mu_0^2e^{2t}=\bar{M}_-^2(t)$ as a renormalization scale and the logarithm of $\bar{s}_2$ is less than $\mathcal{O}(1)$, we find that the RG improved effective potential up to $L$-th-to-leading log order can be calculated by $L$-loop effective potential and $(L+1)$-loop $\beta$ and $\gamma$ functions.
In section~\ref{sec3} and \ref{sec4}, we solve the $\mu_0^2e^{2t}=\bar{M}_-^2(t)$ with respect to $\phi$.
This means that the $\phi$ is not a variable of the effective potential but becomes a fucntion of $\beta$ and $t$.
By using the $\phi$ we can evaluate the mass eigenvalue $\bar{M}_+^2$ and the RG improved effecive potential.
Then, we examine if the logarithm of the ratio of $\bar{M}_+(t)^2$ to $\bar{M}_-(t)^2$ satisfies $\bar{s}_2\lesssim\mathcal{O}(\bar{\lambda}_1)$.
If it is satisfied, the RG improved effective potential can be obatained as mentioned above.
On the other hand, if $\bar{s}_2>\mathcal{O}(\bar{\lambda}_1)$, the heavy particle should be decoupled.
In section~\ref{sec4}, we study such a situation.
We absorb the large logarithm into the new parameters defined in low-energy scale and derive the RG equations described in terms of the redefined parameters.
And then, the RG improved effective potential can be constructed in the low-energy region.
\par

There are three features in this method.
First, we don't need to change the choice of the renormalization scale beyond the leading log order.
This is because since we analyze the logarithmic structure of the effective potential at any  loop order, the choice $\mu_0^2e^{2t}=\bar{M}_-(t)^2$ is valid for the RG improvement up to arbitrary $l$-th-to-leading log order.
Due to this, the $\phi$ which satisfies $\mu_0^2e^{2t}=\bar{M}_-^2(t)$ is the same as the one in the leading log oder.
So we don't need to resolve $\mu_0^2e^{2t}=\bar{M}_-(t)^2$ with respect to $\phi$.
Note that the RG equations must be solved in a loop level corresponding to the desired leading log order.
Second, we can derive the RG improved effective potential without introducing multiple renormalization scales or a step funtion by which the heavy particle is automatically decoupled.
Third, we can decouple the heavy particle from the theory by expanding the quantum correction to the effective potential with respect to $\phi^2/m^2$.
If the logarithm $\log\left(\phi^2/m^2\right)$ is absorbed into the parameters in the low-energy scale, we can derive the RG improved effective potential.\par

Our method can be applied to other multiple scalar model.
If muliple scalar fields are introduced in a model, one represents the classical background fields in terms of polar coordinate such as $(\phi_1,\phi_2)=(\phi\cos\beta,\phi\sin\beta)$.
With $\mu_0^2e^{2t}=\bar{M}_{\text{lightest}}(t)^2$ chosen as a renormalization scale, the $\phi$ corresponding to a radius of the polar coordinate becomes a funcion of the renormalization scale $t$ and angles in the polar coordinate apart from whether it can be solved analytically.
If one attains to this stage, one can implement the calculation of the RG improved effective potential in the same way as this paper.
Finally, since the stability issue or the origin of spontaneous symmetry breaking are investigated through the RG improved effective potential, our work contributes to such studies in a multiple scalar theory.

\section*{Acknowledgement}
We thank T. Morozumi and Y. Shimizu for reading our manuscript and giving useful comments.


\appendix
\section{$\beta$ and $\gamma$ functions in two real scalar model}\label{app1}
In this appendix, we provide the $\beta$ and $\gamma$ functions in a two real single scalar model,
\begin{align}
\begin{split}\nonumber
\beta_{\lambda_1}&=\frac{3}{16\pi^2}(\lambda_1^2+\lambda_3^2),\\
\beta_{\lambda_2}&=\frac{3}{16\pi^2}(\lambda_2^2+\lambda_3^2),\\
\beta_{\lambda_3}&=\frac{\lambda_3}{16\pi^2}(\lambda_1+\lambda_2+4\lambda_3),\\
\gamma_{m_1^2}&=-\frac{1}{16\pi^2m_1^2}(\lambda_1m_1^2+\lambda_3m_2^2),\\
\gamma_{m_2^2}&=-\frac{1}{16\pi^2m_2^2}(\lambda_2m_2^2+\lambda_3m_1^2),\\
\gamma_\Lambda&=-\frac{1}{32\pi^2\Lambda}(m_1^4+m_2^4)\\
\gamma_{\phi_1}&=0,\\
\gamma_{\phi_2}&=0.
\end{split}
\end{align}

\end{document}